\documentclass[aps,twocolumn,prl,superscriptaddress,amsmath,amssymb,showpacs]{revtex4-1}
\pdfoutput=1

\usepackage[breaklinks=true,colorlinks,citecolor=blue,linkcolor=blue,urlcolor=blue]{hyperref}
\usepackage{graphicx}
\usepackage{bm}
\usepackage{ulem}
\usepackage{cancel}
\usepackage{bbold}
\usepackage{color}
\usepackage{braket}

\newcommand\lp{\left(}
\newcommand\rp{\right)}
\newcommand\lsp{\left[}
\newcommand\rsp{\right]}
\newcommand\Imag{\Im m}
\newcommand\Tr{{\rm Tr}}
\newcommand\us{\uparrow}
\newcommand\ds{\downarrow}

\begin{document}

\title{Topological pumping in class-D superconducting wires}

\author{Marco Gibertini}
\email{marco.gibertini@epfl.ch}
\thanks{Present address: Theory and Simulation of Materials (THEOS), \'Ecole Polytechnique F\'ed\'erale de Lausanne, CH-1015 Lausanne, Switzerland.}
\affiliation{NEST, Scuola Normale Superiore and Istituto Nanoscienze-CNR, I-56126 Pisa, Italy}
\author{Rosario Fazio}
\affiliation{NEST, Scuola Normale Superiore and Istituto Nanoscienze-CNR, I-56126 Pisa, Italy}
\author{Marco Polini}
\affiliation{NEST, Istituto Nanoscienze-CNR and Scuola Normale Superiore, I-56126 Pisa, Italy}
\author{Fabio Taddei}
\affiliation{NEST, Istituto Nanoscienze-CNR and Scuola Normale Superiore, I-56126 Pisa, Italy}

\date{\today}

\begin{abstract}
We study adiabatic pumping at a normal metal/class-D superconductor hybrid interface when superconductivity is induced through the proximity effect in a spin-orbit coupled nanowire in the presence of a tilted Zeeman field. When the induced order parameter in the nanowire is non-uniform, the phase diagram has isolated trivial regions surrounded by topological ones. We show that in this case the pumped charge is quantized in units of the elementary charge $e$ and has a topological nature.
\end{abstract}
\maketitle

{\it Introduction.}--- In a quantum system, the periodic modulation of two (or more) independent parameters may give rise to a dc flow of charge in 
the absence of any applied bias voltage. This transport mechanism is known as charge pumping~\cite{Thouless1983,Brouwer1998,Zhou1999,Giazotto2011}. 
When the  period $T$ of the modulation is much larger than any characteristic time scale of the system, pumping is adiabatic and it is related to the 
Berry phase accumulated during the cyclic evolution. For a system coupled to leads the pumped charge can be expressed (ignoring electron-electron interactions) in 
terms of the associated scattering matrix~\cite{Buttiker1994,Brouwer1998}. 

An interesting case, also in view of its metrological applications,  is when the pumped charge is quantized (see e.g.~\cite{quantpump}).  
In these cases a careful estimation of any possible deviation from the quantized value is very important.  If, however,  quantisation stems from topology, 
pumping will be immune from errors and robust to unavoidable perturbations (the recent discovery of topological insulators has also lead to a topological
classification for spin pumps~\cite{meidan10}). 

\begin{figure}
\includegraphics[width=\columnwidth,clip=true,trim=0 4.5mm 0 7.5mm]{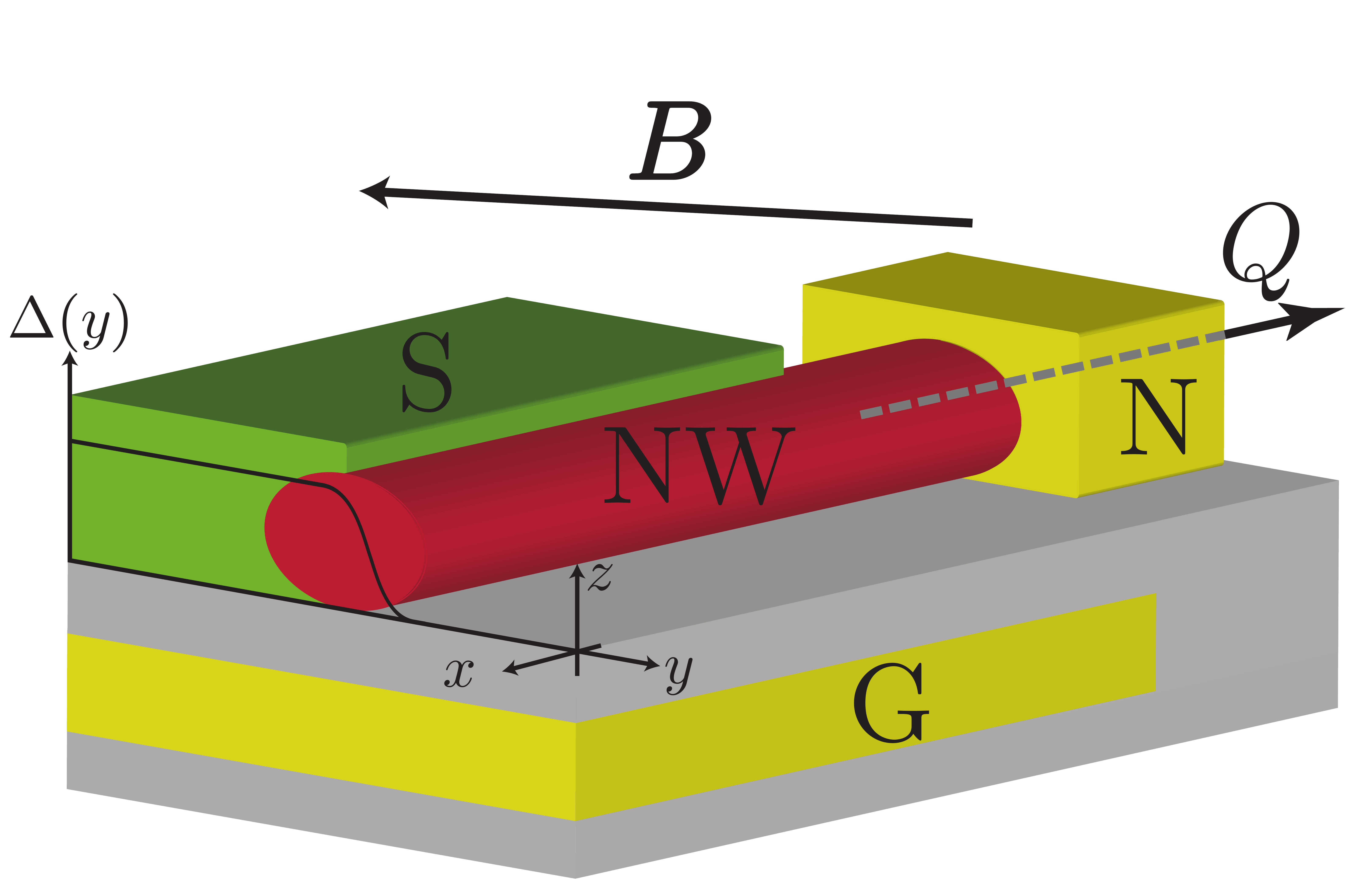}
\caption{(color online) Sketch of the system. A spin-orbit coupled  nanowire (NW) is subject to a tilted Zeeman field $\bm{B}$ and in 
contact with a bulk superconductor (S) which induces a pairing gap $\Delta$. Due to the lateral superconducting contact, the induced gap changes along the transverse direction $y$.  On one side, the wire is attached to a normal lead (N).
A charge $Q$ is pumped through the NW-N interface by changing in time the chemical potential (for example applying a time-dependent voltage to the gate G) and the Zeeman field.
\label{fig:setup}}
\end{figure}

The experimental realization of a topological charge pump is  thus of paramount importance both in fundamental and applied science.
In his pioneering work, Thouless~\cite{Thouless1983} showed that quantized adiabatic pumping (AP) can occur through one-dimensional insulating 
systems as a consequence of the topological properties of the Hamiltonian, strictly related to the existence of gapless points enclosed by the 
adiabatic pumping cycle in parameter space~\cite{Simon1983}. The same topological argument holds even for a finite system attached to 
leads~\cite{Onoda2006}. 
In this work we find that topological charge pumping can be realized in systems which are now intensively investigated both from a
theoretical and experimental point of view, namely normal metal/class-D superconductor hybrid systems.  
As sketched in Fig.\ref{fig:setup},  we consider a spin-orbit coupled semiconductor nanowire in proximity to a $s$-wave superconductor, subject to a Zeeman field and attached to a normal lead on one end.
This is the  same setup, proposed in~\cite{Snano},  that has lead  to  the first evidence~\cite{Mourik2012} of Majorana 
modes in condensed matter~\cite{nanowire}.

When time-reversal symmetry is broken ({\it i.e.} the superconductor is in symmetry class D~\cite{topclass}) the topological classification 
distinguishes two possible phases of the superconductor~\cite{topins,nanowire}, depending on whether 
the system supports or not mid-gap, Majorana, states.  In Refs.~\cite{Lutchyn2011,Stanescu2011,Gibertini2012} it has been shown that when the Zeeman 
field is aligned parallel to the wire (along, {\it e.g.}, the $x$-direction), the phase diagram as a function of chemical potential $\mu$ and  Zeeman field $B_x$ can be divided into simply-connected trivial/topological regions.
In particular, since the proximity effect is induced by contact with a bulk superconductor from one side of the nanowire~\cite{Mourik2012,Das2012,Deng2012,Finck2013}, the induced gap is not expected to penetrate uniformly in the direction transverse to the wire axis.
It rather decreases when one moves away from the side in contact with the superconductor.
As it will be shown in the following, the concomitant presence of a space-dependent profile of the induced gap and a transverse component of the Zeeman field has profound consequences on the phase diagram that are crucial to the realization of topological pumping in these systems.

\begin{figure}
\includegraphics{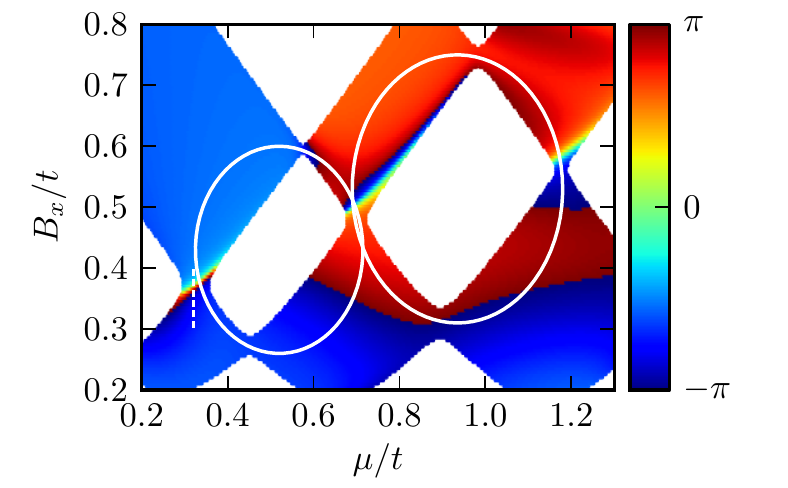}
\caption{(color online)  Color plot of the phase of the Andreev reflection amplitude $r_{he}$ as a function of the longitudinal Zeeman field $B_x$ and the 
chemical potential $\mu$ in the superconducting region. Inside colored regions the superconductor is in the topologically  non-trivial phase while inside 
white regions $r_{he}=0$ and the superconductor is in the trivial phase. These results have been obtained with the following set of parameters: $W/a = 10$, 
$\alpha/t=0.26$, and $B_z/t=0.08$, where $a$ is the lattice constant. Inside the normal lead $\mu = B_x = 0$ so that only a  single mode is propagating.
\label{fig:singlemode}}
\end{figure}

{\it Model of the nanowire.}--- 
The Hamiltonian of a nanowire in proximity with a superconductor (see Fig.\ref{fig:setup}) reads
\begin{equation}
\hat {\cal H} = \sum_{i,j,\sigma,\sigma'} h_{ij, \sigma,\sigma'}\hat c^\dag_{i,\sigma}\hat c_{j,\sigma'}
+ \sum_{ i} [\Delta(y_i)~\hat c^\dag_{i,\us}\hat c^\dag_{i,\ds} + \text{H.c.}]
\label{ham}
\end{equation}
with 
\begin{equation}
h_{ij} = [ (4t-\mu)\,\mathbb{1} + {\bm B} \cdot {\bm \sigma }]\,\delta_{ij} \nonumber\\
-[ t\,\mathbb{1} -i\alpha(  \nu'_{ij}\sigma^x - \nu_{ij}\sigma^y)]\,\delta_{\langle i,j\rangle}~,
\end{equation}
where $c^\dag_{i,\sigma}$ ($c_{i,\sigma}$) creates (destroys) an electron at site $i$ with spin $\sigma$, $\delta_{\langle i,j\rangle}$ restricts sites
$i$ and $j$ to be nearest-neighbors, $\sigma^{b}$ (with $b=x,y,z$) are spin-1/2 Pauli matrices, $\nu_{i j} = \hat{\bm x} \cdot \hat {\bm d}_{i j}$, and 
$\nu'_{i j} = \hat{\bm y} \cdot \hat {\bm d}_{i j}$, with $\hat {\bm d}_{i j}$ being the unit vector connecting 
site $j$ to site $i$. In Eq.~(\ref{ham}) $t$ is the hopping energy, $\alpha$ the spin-orbit coupling  strength, and  $\mu$ the chemical potential. 
The only non-zero components of the Zeeman field are $B_x$ and $B_z$. We adopted a tight-binding formulation as this will be used in the numerical simulations. The pairing gap $\Delta$ varies along the $y$-direction (with $0\leq y \leq W$, $W$  being the width of the wire) because of the lateral contact to the bulk superconductor.
We have considered different profiles of the pairing amplitude along the transverse direction~\cite{footnote}, but the picture we are going to illustrate does not depend (to a large extent) on these details.
The results presented here refer to a semi-infinite nanowire coupled to a normal electrode on one end.
(We have found analogous results for a nanowire of finite length, provided that the latter is long enough so that the coupling between a Majorana fermion and the normal lead is larger than the coupling between the two Majorana fermions at the ends of the nanowire.)
The phase diagram of the model in Eq.~\eqref{ham} is shown in Fig.~\ref{fig:singlemode}. The most important feature to be noted is that the topologically trivial regions  (in white) form isolated ``islands'' and the phase diagram is multiply connected.
This is one of the two factors enabling topological pumping in our setup.

Since quasiparticle transport is strongly suppressed in a sufficiently long superconductor 
due to the gap, the charge $Q$ pumped through the normal lead is obtained from the evolution in time of the normal ($r_{ee}$) and Andreev ($r_{he}$) 
components of the reflection matrix $r$ of the hybrid system~\cite{pumpedSC}:
\begin{equation}\label{eq:blaauboer}
Q = \frac{e}{2\pi}\int_{0}^{T} \Imag\Tr\left[ \frac{d r_{ee}(\tau)}{d\tau} r^\dag_{ee}(\tau) -  \frac{d r_{he}(\tau)}{d\tau} r^\dag_{he}(\tau)\right] d\tau~.
\end{equation}
Adiabaticity is achieved when $T$ is much larger than the quasiparticle dwell time inside the superconductor.

In what follows we first analyze the case of a single-mode lead, showing that the pumped charge is topological and quantized in units of the elementary charge $e$.
We then show that this behavior can be understood from simple arguments.
Finally, we then demonstrate that topological AP can be obtained, in a (topological) superconductor, 
even relaxing the constraint of a single open channel in the lead by using a quantum point contact (QPC) to couple the lead to the superconductor.
Moreover, our results are robust against disorder.

{\it Pumping in a single-mode lead.}---We first consider the situation in which the normal lead coupled to the superconducting nanowire supports a single open channel.
The topological invariant ${\cal N}$ distinguishing topologically trivial from non-trivial phases is related to the reflection matrix $r$ at the normal-metal/superconductor (NS) interface according to ${\cal N}=\det\,r$~\cite{Akhmerov2011}. For a single-mode normal lead this reduces to ${\cal N} = |r_{ee}|^2-|r_{he}|^2$.
When the superconductor is in the topological phase (${\cal N}=-1$), normal reflection is absent, $r_{ee}=0$, and the mode is perfectly Andreev-reflected by the Majorana 
fermion at the NS boundary, $|r_{he}| = 1$~\cite{Wimmer2011,Beri2009}. On the contrary, when the superconductor is in the trivial phase (${\cal N}=+1$), the mode is fully reflected ($|r_{ee}|=1$). We have evaluated the  reflection matrix by using a method which combines recursive Green's function algorithms~\cite{Sanvito1999} with a wave-function matching approach~\cite{Zwierzycki2008}. In Fig.~\ref{fig:singlemode} the superconductor is in the non-trivial phase ($|r_{he}| =1$) inside colored regions and the color coding depicts the phase of the Andreev reflection 
amplitude as a function of $B_x$ and $\mu$ in the superconducting region. 
Inside the white regions the superconductor is in the topologically trivial phase and $r_{he}$ vanishes. At the boundary of such white regions the gap closes, signaling a quantum phase transition between topologically trivial and non-trivial phases.  
As a consequence, the topological sector of the phase diagram is non-simply connected and the elliptic paths marked by white solid lines in Fig.~\ref{fig:singlemode} cannot be shrunk to a point without crossing the phase boundary.
As we shall see in details below, it is the phase of the Andreev reflection amplitude $r_{he}$ that plays a crucial role in pumping.
\begin{figure}
\includegraphics{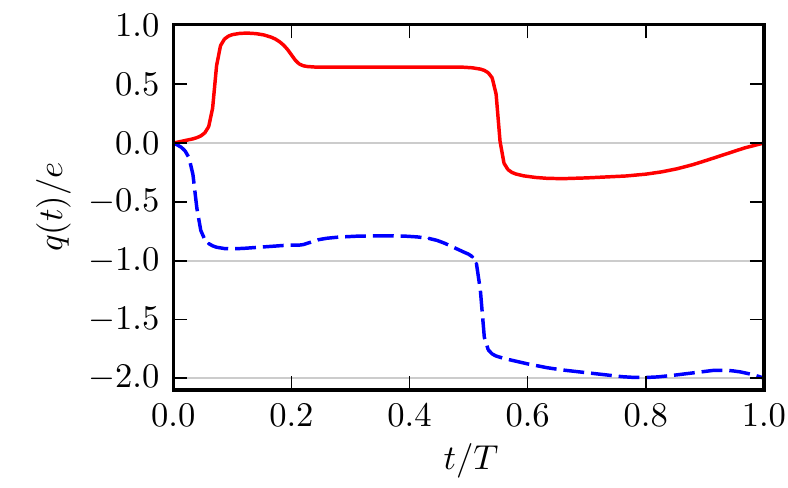}
\caption{(color online) Cumulative pumped charge $q(t)$ (in units of $e$) as a function of time $t$ (in units of the pumping period $T$). The thick solid line refers to the leftmost elliptic path in Fig.~\ref{fig:singlemode}, while the thick dashed line to the rightmost one. At the end of the cycle, the final pumped charge $Q \equiv q(T)$ is quantized.
\label{fig:singlemodepump}}
\end{figure}

We have computed the adiabatic pumped charge along both paths shown in Fig.~\ref{fig:singlemode} 
by using a gauge-invariant discretized form of Eq.~\eqref{eq:blaauboer}. We modeled the evolution in time of the pumping parameters as follows
\begin{equation}\label{eq:adparameter}
\left\{
\begin{aligned}
&\mu(t)  = \bar\mu + \delta\mu \cos\lp 2\pi\, t/T\rp \\
&B_x(t) = \bar B_x + \delta B_x \sin\lp 2\pi\, t/T\rp\\
\end{aligned}
\right.~,
\end{equation} 
where $\bar \mu/t=0.52$, $\delta\mu/t=0.195$, $\bar B_x/t=0.43$, and  $\delta B_x/t=0.17$ for the left path, while $\bar \mu/t=0.935$, $\delta\mu/t=0.245$, $\bar B_x/t=0.53$, and  $\delta B_x/t=0.22$ for the right one. 
Experimentally, this evolution could be realized by properly designing metal gates and coils to introduce time-dependent gate voltages and magnetic fields acting only on the superconducting section of the nanowire.  

In Fig.~\ref{fig:singlemodepump} we show our numerical results for the cumulative pumped charge $q(t)$ defined as in Eq.~\eqref{eq:blaauboer} with the integral running from 0 to $t$. The solid (dashed) line refers to the leftmost (rightmost) path in Fig.~\ref{fig:singlemode}.
The total pumped charge in a cycle $Q \equiv q(T)$ is strictly quantized and equal to $0$ and $-2\,e$, respectively.
It should be clear that the value of the pumped charge is topological, {\it i.e.} it does not change under smooth deformations of the pumping path in parameter space.
We have verified that topological pumping occurs in different parameter ranges or with different profiles of the pairing amplitude.
We have noticed that $Q/e$ can assume any integer value, even or odd, depending on the details of the system under consideration.

{\it Origin of the topological pumping.} --- We shall now show that the presence of a single transmitting channel at the interface between the normal electrode and the superconducting wire is the second key to obtain topological pumping.
In this case the pumped charge can be expressed in an elegant simple form.
When the superconductor is in symmetry class D and the normal lead supports a single mode, the reflection matrix reduces to a single amplitude, {\it i.e.} a quasiparticle can undergo either normal or  Andreev reflection.
Unitarity of the scattering matrix requires that the corresponding amplitude is a phase factor $e^{i\beta}$~\cite{Beri2009,Wimmer2011}. The pumped charge is then given by
\begin{equation}
\label{ne}
Q = \pm\frac{e}{2\pi} \int_{0}^{T} \frac{d \beta}{d t}~ dt =\pm e\frac{\beta(T)-\beta(0)}{2\pi} = \pm n e~,
\end{equation}
where the upper (lower) sign is for normal (Andreev) reflection and $n$ is an integer since scattering amplitudes must be single-valued.

If the phase diagram in parameter space is simply connected and the pumping path can be continuously contracted to a point, the integer $n$ must be zero. 
On the contrary, if the phase diagram is non-simply connected and the pumping path, running in the topological phase, encloses a non-topological region, no restriction is enforced on $n$. Therefore the adiabatically pumped charge can, in principle, be topological, in the sense that any continuous deformation of the pumping path in parameter space does not change the integer $n$~\cite{quantization}.
As shown above, $n$ is in general different from zero for the setup described in Fig.~\ref{fig:setup}, thus proving that topological pumping of finite charge does occur for such systems.

From Fig.~\ref{fig:singlemode}, we note that the phase of $r_{he}$ is nearly constant throughout parameter space with the exception of ``stripes'' connecting different non-topological (white) regions, across which the phase winds by approximately $2\pi$ ~\cite{SM}.
When the pumping path coincides with non-contractible elliptic paths of the type shown in Fig.~\ref{fig:singlemode}, these resonant stripes, according to Eq.~(\ref{ne}), contribute by roughly $e$ to the pumped charge $Q$.
The topological nature of pumping stems from the fact that it is not possible to deform the contour in order to avoid the crossing of the resonant stripes without intersecting the phase boundary lines.

\begin{figure}
\includegraphics{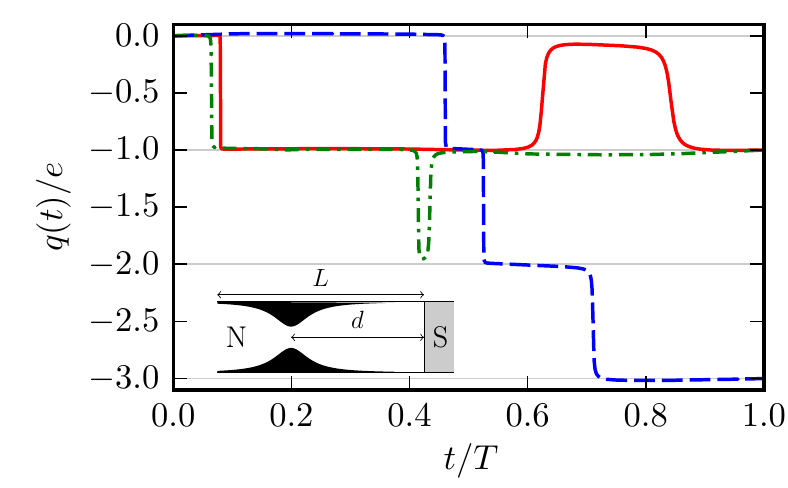}
\caption{(color online) Cumulative pumped charge $q(t)$ (in units of $e$) as a function of time $t$ (in units of the pumping period $T$) in the presence of a QPC at a distance $d=60~a$ from the NS interface. At the end of the cycle the final pumped charge $Q \equiv q(T)$ is quantized.
Results for different trajectories in parameter space are denoted with distinct line styles.  The following set of parameters has been adopted: $W/a=10$,  $\alpha/t=0.15$ and $B_z/t=0.1$. In the inset we have sketched the setup with a QPC at a distance $d$ from the NS boundary in order to restrict the number of propagating modes. In simulations with disorder, random potential fluctuations are considered in a region $L>d$ inside the normal lead before the NS interface.
\label{fig:qpcpump}}
\end{figure}

{\it QPC geometry.}---We now show that topological AP can be achieved in a more realistic situation in which the number of open channels in the normal electrode is arbitrary (so that $r_{ee}$ and $r_{he}$ become matrices) but a QPC restricts the number of modes incident on the NS boundary (see inset in Fig.~\ref{fig:qpcpump}).  The QPC consists of a saddle-point constriction~\cite{Buttiker1990,Wimmer2011} at a distance $d$ from the NS interface (see Fig.~\ref{fig:qpcpump}). 
When the pinch-off potential of the QPC is chosen so that only a single mode is transmitted, it is possible to prove (we assume the superconductor to be in the topological phase) that the matrix $r_{he}$ changes in time only through a global phase factor while $r_{ee}$ remains constant, so that the pumped charge in a cycle is still quantized~\cite{SM}. 

This is confirmed by the numerical results in Fig.~\ref{fig:qpcpump}, which show the charge pumped along three different non-contractible paths.
The pumping parameters $B_x$ and $\mu$ in the superconductor follow the time evolution in Eq.~\eqref{eq:adparameter}, while in the normal lead we fix $B_x\equiv \bar B_x$ and $\mu\equiv\bar\mu$.
The three curves in Fig.~\ref{fig:qpcpump} simply differ in the values of $\bar B_x$, $\delta B_x$, $\bar\mu$, and $\delta\mu$, 
which identify the pumping paths through Eq.~\eqref{eq:adparameter}. 

The pumped charge $Q$ is quantized to integer multiples of $e$. As in the case of a single-mode lead, the value of $Q$ can be interpreted as the result of contributions from the crossing of resonant stripes which gives rise to a twist of the global phase of $r_{he}$ by $\sim 2\pi$ (data not shown). Each time the system crosses a resonant stripe approximately a charge $e$ is pumped in the normal lead, corresponding to the step-like features in Fig.~\ref{fig:qpcpump}. When the distance $d$ is varied, the interference pattern changes in such a way that resonant stripes could come in/out of the way of the pumping path. We have noticed that, although the value of $Q/e$ can vary, it is always strictly quantized to integer values and the parity of $Q/e$ is conserved. 

\begin{figure}
\includegraphics{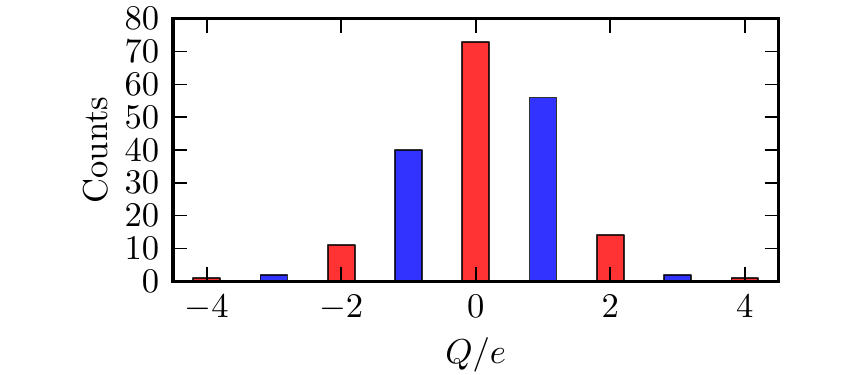}
\caption{(color online) Histogram of the pumped charge $Q/e$ obtained for 100 disorder realizations with $L/a=200$, $d/a=140$, and $U_{\rm dis}/t=0.5$. Light gray (red) and dark gray (blue) bars refer to two different sets of parameters and distinct pumping paths. We notice that although the value of $Q$ depends on the disorder configuration, it is always quantized and its parity is preserved.  \label{fig:vsdisorder}}
\end{figure}

We have also investigated the effect of disorder in the QPC by introducing a random on-site potential in a region of length $L$ ($>d$) on the left of the NS interface (see Fig.~\ref{fig:qpcpump}). Fluctuations of the on-site energy lie in the interval $[-U_{\rm dis}/2,U_{\rm dis}/2]$. 
We have noticed that quantization of the pumped charge is almost exactly preserved up to large values of the disorder strength ($U_{\rm dis}/t\approx 5-6$) provided that a single mode is transmitted through the QPC (see Fig.~\ref{fig:qpcpump}). As shown in Fig.~\ref{fig:vsdisorder}, the details of the specific disorder configuration affect the value of the pumped charge but, given the system parameters and the pumping path, the parity of $Q/e$ is insensitive to smooth changes of the Hamiltonian.

%
{\it Conclusions} --
We have shown that topological adiabatic pumping occurs in a class-D superconducting nanowire connected to a metallic lead on one end. 
This happens when the lead supports a single propagating mode or the nanowire is coupled to the lead through a quantum point contact. 
The necessary condition to achieve a topological pumped charge is that the phase diagram presents a non-simply connected 
structure, where isolated non-topological regions are surrounded by connected topological ones. This is possible by allowing both a non-uniform pairing 
amplitude and a tilted Zeeman field. Non-contractible pumping paths in parameter space can thus be identified within the topological phase.

{\it Acknowledgements} -- 
We would like to acknowledge fruitful discussions with C.W.J. Beenakker. 
This work has been supported by the EU FP7 Programme under Grant Agreement  
No. 234970-NANOCTM, No. 248629-SOLID, No. 233992-QNEMS, No. 238345-GEOMDISS, and No. 215368-SEMISPINNET.


\begin{appendix}
\section{SUPPLEMENTAL MATERIAL}\label{app:SM}
This section contains technical details and numerical results relevant to the main text.

\setcounter{figure}{0}
\setcounter{equation}{0}
\renewcommand{\theequation}{S\arabic{equation}}
\renewcommand{\thefigure}{S\arabic{figure}}

\section{Resonant stripes in the phase of Andreev reflection}
According to Fig.~\ref{fig:singlemode} of the main text, the phase of $r_{he}$ is almost constant throughout the parameter space with the exception of stripes connecting different non-topological (white) regions, across which the phase winds by approximately $2\pi$.
When the pumping path coincides with non-contractible elliptic paths of the type shown in Fig.~\ref{fig:singlemode}, these resonant stripes, according to Eq.~\eqref{ne} in the main text, contribute by roughly $e$ to the pumped charge $Q$.
Along the rightmost path both resonant stripes give rise to a counterclockwise rotation of the phase of $r_{he}$  so that their corresponding contributions sum up to $-2 e$. Along the leftmost path the winding of the Andreev reflection amplitude phase across the resonant stripes is opposite and the total pumped charge vanishes.

\begin{figure}
\includegraphics{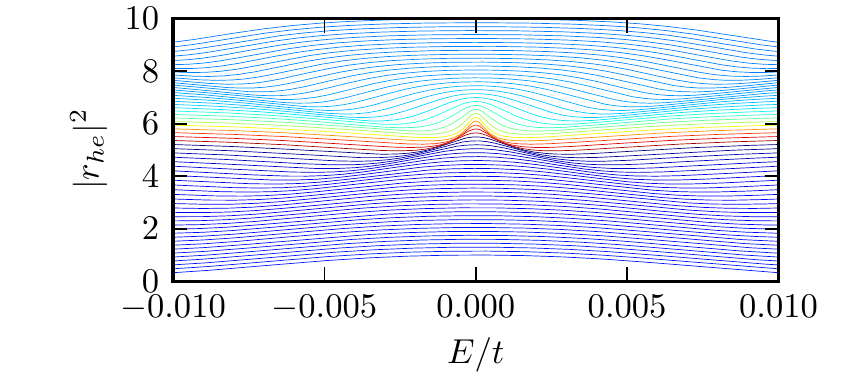}
\caption{(color online) Andreev reflection $|r_{he}|^2$ as a function of energy $E$ for several increasing values of the longitudinal Zeeman field, $0.3\leq V_x/t\leq 0.42$, at fixed $\mu/t = 0.32$ (see dashed line in Fig.~2 of the main text). Different curves have been offset for clarity. The color coding is associated with the phase of $r_{he}$ at zero energy as in Fig.~\ref{fig:singlemode}. 
\label{fig:resonant}}
\end{figure}

The scattering approach we are using does not allow to address the direct role of Majorana fermions in the pumping mechanism.
Nevertheless, we notice that the resonant stripes are associated with a narrowing of the Majorana-induced zero-energy resonance in the Andreev reflection, distinctive of the non-trivial phase~\cite{Law2009,Flensberg2010}. This is shown in Fig.~\ref{fig:resonant} where we plot $|r_{he}|^2$ (proportional to the differential conductance of the NS interface) as  a function of energy. Different curves are reported (and offset for clarity) for several values of $V_x$ at fixed $\mu/t = 0.32$, corresponding to points along the dashed segment in Fig.~2.

\section{Pumped charge quantization in the presence of a QPC}
In this section we shall prove that when the QPC pinch-off potential is such that a single mode is completely transmitted, the pumped charge is still given by Eq.~\eqref{ne} of the main text, and therefore is quantized.
We write the scattering matrix of a symmetric QPC as
\begin{equation}\label{eq:Sqpc}
s_{\rm QPC} = \begin{pmatrix}
r & t \\
t & r
\end{pmatrix}~.
\end{equation}
The ``orthogonality'' relation between the reflection and transmission sub-blocks
\begin{equation}\label{eq:orthog}
r^{\dag} t = 0
\end{equation}
ensures that each mode is either completely transmitted or reflected.
Indeed, making use of the unitarity of the scattering matrix $s_{\rm QPC}$, we find
\begin{equation}
\label{ses}
0 = (r^\dag t)^{\dag} r^\dag t=  t^{\dag} \lp \openone - t t^{\dag} \rp t = t^{\dag} t  \lp \openone -  t^{\dag} t  \rp~,
\end{equation}
yielding that the eigenvalues $T_{n}$ of the Hermitian matrix $ t^{\dag} t $ are either $0$ or  $1$.
In addition, from (\ref{ses}), we have that $ t^{\dag} t = (t^{\dag} t)^2 $, {\it i.e.} the matrix $t^{\dag} t$ is a projector on the subspace spanned by the modes which are trasmitted. 

In particular, when the QPC is open  only for a single mode, $T_n = \delta_{n,1}$, we have that $t^{\dag} t \equiv \ket{1}\bra{1}$, $\ket{n}\bra{n}$ being a projector on the $n$-th eigenstate of $t^{\dag} t$.
The last equality is satisfied by setting
\begin{align}\label{eq:tmodebasis}
t  = \sum_{n} \chi_{n}\ket{n}\bra{1}
\qquad\text{with}\qquad
\sum_{n}\left\vert\chi_n\right\vert^2 = 1~.
\end{align}
By combining the scattering matrix of the QPC in Eq.~\eqref{eq:Sqpc} with the Andreev scattering matrix $s_{\rm A}$ at the NS interface~\cite{Beenakker1992} we obtain the normal reflection matrix 
\begin{equation}\label{eq:generalrhe}
r_{ee} =r +  t B t~,
\end{equation}
and the Andreev reflection matrix 
\begin{equation}\label{eq:generalree}
r_{he} = t^* A t~,
\end{equation}
where $A$ and $B$ are in general complicated operators~\cite{cleanNS}.
Unitarity of the scattering matrix requires that 
\begin{equation}\label{eq:unitarity}
r_{he}^{\rm t}r_{ee} + r_{ee}^{\rm t}r_{he} = 0~.
\end{equation}
From equations \eqref{eq:tmodebasis}-\eqref{eq:generalrhe} we find that 
\begin{align}
r^{\rm t}_{he}r_{ee}  &= t^{\rm t} A^{\rm t} t^\dag \lp r + t B t\rp = t^{\rm t} A^{\rm t} \ket{1}\bra{1} B t \nonumber\\
&= \ket{1}\bra{1}\lp\sum_{m}\chi_{m}\bra{1}A\ket{m}\rp\lp\sum_{n}\chi_{n}\bra{1}B\ket{n}\rp\nonumber\\
&= t^{\rm t} B^{\rm t} \ket{1}\bra{1} A t  =\lp r^{\rm t} + t^{\rm t} B^{\rm t} t^{\rm t}\rp t^* A t \nonumber\\
&= r_{ee}^{\rm t}r_{he}
\end{align}
so that Eq.~\eqref{eq:unitarity} leads to $r^{\rm t}_{he}r_{ee}=0$. This means that each mode is either completely reflected or Andreev reflected. Indeed, exploiting the unitarity and particle-hole symmetry of the reflection matrix, we have
\begin{align}
0 &= r^\dag_{he} r^*_{ee} (r^\dag_{he} r^*_{ee})^\dag  = r^\dag_{he} (\openone - r_{eh}r^\dag_{eh})^* r_{he} \nonumber\\
&= r^\dag_{he} (\openone - r_{he}r^\dag_{he}) r_{he} = r^\dag_{he} r_{he}(\openone - r^\dag_{he} r_{he})~.
\end{align}
In addition the condition $r^{\rm t}_{he}r_{ee}=0$ requires that
\begin{equation}\label{eq:condition}
\lp\sum_{m}\chi_{m}\bra{1}A\ket{m}\rp\lp\sum_{n}\chi_{n}\bra{1}B\ket{n}\rp = 0~. 
\end{equation}
As we shall see below the first factor must be different from zero when the superconductor is in the topologically non-trivial phase and consequently we need to have
\begin{equation}
\sum_{n}\chi_{n}\bra{1}B\ket{n} = 0~. 
\end{equation}
Thus, the normal reflection matrix coincides with the reflection matrix of the QPC. Indeed
\begin{align}
r_{ee} &= r  + tBt \nonumber\\
&= r + \sum_{m}\chi_{m}\ket{m}\bra{1} \lp\sum_{n}\chi_{n}\bra{1}B\ket{n}\rp = r~.
\end{align}
As a consequence $r_{ee}$ does not depend on time and does not contribute to the pumped charge.

As far as the Andreev reflection matrix is concerned, with the help of Eq.~\eqref{eq:tmodebasis} we find
\begin{align}\label{eq:rhesinglemode}
r_{he}  = \sum_{n} \psi_{n}\ket{n} \bra{1}~,
\end{align}
with
\begin{align}
\psi_{n} = \chi_{n} \lsp \sum_{m}\chi_{m}~\bra{1} A \ket{m}\rsp~,
\end{align}
so that 
\begin{align}
r^{\dag}_{he}r_{he}  = \lsp\sum_{n} \left|\psi_n\right|^2 \rsp \ket{1}\bra{1}~.
\end{align}
From the discussion above we know that $\sum_n\left|\psi_n\right|^2$ is either exactly $0$ or $1$. On the other hand, when the superconductor is in the topological phase the Andreev reflection matrix can not vanish~\cite{firstfactor} and thus the Hermitian matrix $r^{\dag}_{he}r_{he}$ must be a projector on the transmitted mode $\ket{1}$ ($\sum_n\left|\psi_n\right|^2=1$). 
As a consequence,  $r_{he}$ defines through Eq.~\eqref{eq:rhesinglemode} a vector $\psi$ with components $\psi_n\in\mathbb{C}$ such that $\psi^{\dag}\psi = 1$. During the adiabatic evolution, the coefficients $\chi_n$ are fixed and $\psi$ can change only through the global phase factor
\begin{equation}
e^{i \varphi} = \sum_{m}\chi_{m}~\bra{1} A \ket{m}~.
\end{equation}
Reminding that the reflection matrix does not depend on time, with the help of Eq.~\eqref{eq:rhesinglemode} the pumped charge is given by
\begin{align}
Q &= -\frac{e}{2\pi} \int_{0}^{T} \Imag \Tr\left[\frac{d r_{he}(t)}{dt} r^\dag_{he}(t)\right] dt \nonumber\\
& = -\frac{e}{2\pi} \int_{0}^{T} \Imag \sum_{n}\psi_n^*\frac{d \psi_n}{dt}dt  \\ 
& =  -\frac{e}{2\pi} \int_{0}^{T} \Imag\lsp\psi^\dag\frac{d \psi}{dt}\rsp dt =  -\frac{e}{2\pi} \int_{0}^{T} \frac{d\varphi}{dt} dt = -e n ~,\nonumber
\end{align}
so that the pumped charge in a cycle is quantized.
\end{appendix}



\begin{thebibliography}{77}	

\bibitem{Brouwer1998}
	P.W.~Brouwer, \href {http://dx.doi.org/10.1103/PhysRevB.58.R10135}{ \prb~\textbf{58}, R10135 (1998)}.

\bibitem{Zhou1999}
	F.~Zhou, B.~Spivak, and B.~Altshuler, \href{http://dx.doi.org/10.1103/PhysRevLett.82.608}{\prl~\textbf{82}, 608 (1999)}.
	
\bibitem{Thouless1983}
	D.J.~Thouless, \href{http://dx.doi.org/10.1103/PhysRevB.27.6083}{\prb~\textbf{27}, 6083 (1983)}.

\bibitem{Giazotto2011}
	F. Giazotto, P. Spathis, S. Roddaro, S. Biswas, F. Taddei, M. Governale, and L. Sorba,
	\href{http://dx.doi.org/10.1038/nphys2053}{Nat. Phys. \textbf{7}, 857 (2011)}.

\bibitem{Buttiker1994}
	M. B\"uttiker, H. Thomas, and A. Pr\^etre, Z. Phys. B \textbf{94}, 133 (1994).

\bibitem{quantpump}
	H. Pothier, P. Lafarge, C. Urbina, D. Esteve and M. H. Devoret, \href{http://iopscience.iop.org/0295-5075/17/3/011}{Europhys.~Lett.~\textbf{17}, 249 (1992)}; 
	J. M. Martinis, M. Nahum, and H. D. Jensen, \href{http://prl.aps.org/abstract/PRL/v72/i6/p904_1}{\prl \textbf{72}, 904 (1994)}; 
	J.P. Pekola, J.J. Vartiainen, M. M\"ott\"onen, O.-P.i Saira, M. Meschke, and D.V. Averin,
	\href{http://www.nature.com/nphys/journal/v4/n2/full/nphys808.html}{Nature Physics \textbf{4}, 120 (2008)}; S. J. Chorley, J. Frake, C. G. Smith, 
	G. A. C. Jones, and M. R. Buitelaar,  \href{http://scitation.aip.org/content/aip/journal/apl/100/14/10.1063/1.3700967}{Appl. Phys. Lett. \textbf{100}, 143104 (2012)}; M. R. Connolly, K. L. Chiu, S. P. Giblin,
	M. Kataoka, J. D. Fletcher,  C. Chua, J.P. Griffiths, G.A.C. Jones, V.I. Fal'ko, C.G. Smith,  and T.J.B.M. Janssen, 
	\href{http://www.nature.com/nnano/journal/v8/n6/full/nnano.2013.73.html}{Nature Nanotechnology \textbf{8},  417 (2013)}.

\bibitem{meidan10}
	D. Meidan, T. Micklitz, and P. W. Brouwer,  
	\href{http://link.aps.org/doi/10.1103/PhysRevB.82.161303}{\prb~\textbf{82}, 161303 (2010)}.

\bibitem{Simon1983}
	B.~Simon, \href {http://dx.doi.org/10.1103/PhysRevLett.51.2167}{ \prl~\textbf{51}, 2167 (1983)}.

\bibitem{Onoda2006}
	S.~Onoda, C.-H.~Chern, S.~Murakami, Y.~Ogimoto, and N.~Nagaosa, 
	\href {http://dx.doi.org/10.1103/PhysRevLett.97.266807}{\prl~\textbf{97}, 266807 (2006)}.

\bibitem{Snano}
	{R.M.~Lutchyn}, {J.D.~Sau}, and {S.~Das~Sarma},  
	\href {http://dx.doi.org/10.1103/PhysRevLett.105.077001}{ \prl~\textbf{105}, 077001 (2010)};
	{Y.~Oreg}, {G.~Refael}, and {F.~von Oppen},  
	\href {http://dx.doi.org/10.1103/PhysRevLett.105.177002}{ \prl~\textbf{105}, 177002 (2010)};

\bibitem{Mourik2012}
	V.~Mourik, K.~Zuo, S.M.~Frolov, S.R.~Plissard, E.P.A.M.~Bakkers, and L.P.~Kouwenhoven.
	\href{http://www.sciencemag.org/content/336/6084/1003.abstract}{Science \textbf{336},  1003 (2012)}.

\bibitem{nanowire}
	J.~Alicea,
	\href{http://stacks.iop.org/0034-4885/75/i=7/a=076501}{Rep.~ Progr.~Phys.~\textbf{75}, 076501 (2012)};
	C.W.J.~Beenakker, 
	\href{http://www.annualreviews.org/doi/abs/10.1146/annurev-conmatphys-030212-184337}{Annu.~Rev.~Condens.~Matter Phys.~\textbf{4}, 113 (2013)};
	M.~Leijnse and K.~Flensberg,
	\href{http://iopscience.iop.org/0268-1242/27/12/124003/}{Semicond. Sci. Technol.~\textbf{27}, 124003 (2012)}.
			
\bibitem{topclass}
	A.~Altland and M.R.~Zirnbauer, \href{http://dx.doi.org/10.1103/PhysRevB.55.1142}{\prb~\textbf{55}, 1142 (1997)};
	A.P.~Schnyder, S.~Ryu, A.~Furusaki, and A.W.W.~Ludwig, 
	\href{http://dx.doi.org/10.1103/PhysRevB.78.195125}{ {\it ibid.}  \textbf{78}, 195125 (2008)};
	A.P.~Schnyder, S.~Ryu, A.~Furusaki, and A.W.W.~Ludwig,
	\href {http://dx.doi.org/10.1063/1.3149481}{AIP Conference Proceedings \textbf{1134}, 10 (2009)};
	A.~Kitaev,
	\href {http://dx.doi.org/10.1063/1.3149495}{ {\it ibid.} \textbf{1134}, 22 (2009)};
	S.~Ryu, A.P.~Schnyder, A.~Furusaki, and A.W.W.~Ludwig,
	\href{http://stacks.iop.org/1367-2630/12/i=6/a=065010}{ New J.~Phys.~ \textbf{12}, 065010 (2010)}.
		
\bibitem{topins}
	M.Z.~Hasan and C.L.~Kane, 
	\href{http://link.aps.org/doi/10.1103/RevModPhys.82.3045}{\rmp~\textbf{82}, 3045 (2010)};
	X.-L.~Qi and S.-C. Zhang, 
	\href{http://link.aps.org/doi/10.1103/RevModPhys.83.1057}{ {\it ibid.} \textbf{83}, 1057 (2011)}.

\bibitem{Lutchyn2011}
	{R.M.~Lutchyn}, {T.D.~Stanescu}, and {S.~Das~Sarma},  
	\href {http://dx.doi.org/10.1103/PhysRevLett.106.127001}{ \prl~\textbf{106}, 127001 (2011)}.

\bibitem{Stanescu2011}
	T.D.~Stanescu, R.M.~Lutchyn, and S.~Das~Sarma,
	\href{http://link.aps.org/doi/10.1103/PhysRevB.84.144522}{\prb~\textbf{84}, 144522 (2011)}.

\bibitem{Gibertini2012}
	M.~Gibertini, F.~Taddei, M.~Polini, and R.~Fazio,
 	\href{http://link.aps.org/doi/10.1103/PhysRevB.85.144525}{\prb~\textbf{85}, 144525 (2012)}.

\bibitem{Das2012}
	A.~Das, Y.~Ronen, Y.~Most, Y.~Oreg, M.~Heiblum, and H.~Shtrikman,
	\href{http://dx.doi.org/10.1038/nphys2479}{Nat. Phys. \textbf{8}, 887 (2012)}.

\bibitem{Deng2012}
	M.T.~Deng, C.L.~Yu, G.Y.~Huang, M.~Larsson, P.~Caroff, and H.Q.~Xu,
	\href{http://pubs.acs.org/doi/abs/10.1021/nl303758w}{Nano Lett. \textbf{12}, 6414 (2012)}.

\bibitem{Finck2013}
	A.D.K.~Finck, D.J.~Van~Harlingen, P.K.~Mohseni, K.~Jung, and X.~Li,  
	\href {http://dx.doi.org/10.1103/PhysRevLett.110.126406}{ \prl~\textbf{110}, 126406 (2013)}.

\bibitem{footnote}
	Although the results presented here do not depend qualitatively on the profile of the pairing amplitude, for definiteness here we considered 
	$$\Delta_{\rm pow}(y) = \Delta_0 \lsp 1-\theta \lp y /W\rp^{10}\rsp$$ 	\
	and 
	$$\Delta_{\rm exp}(y) = \Delta_0 \left\{ 1-\frac{\theta}{2} \bigg[ 1 + \tanh\lp8 (y/W-1/2)\rp\bigg]\right\}~,$$ 
	where $0\leq\theta\leq 1$ is a dimensionless parameter which measures the suppression of the superconducting pairing across the wire~\cite{Lutchyn2011,Stanescu2011}.
	
\bibitem{pumpedSC}
	J.~Wang, Y.~Wei, B.~Wang, and H.~Guo, \href {http://dx.doi.org/10.1063/1.1421236}{ Appl. Phys. Lett. \textbf{79}, 3977 (2001)};
	{M.~Blaauboer},  \href {http://dx.doi.org/10.1103/PhysRevB.65.235318}{ \prb~\textbf{65}, 235318 (2002)};
	S.~Pilgram, H.~Schomerus, A.~M.~Martin, and  M.~B\"uttiker,  
	\href {http://dx.doi.org/10.1103/PhysRevB.65.045321}{{\it ibid.} \textbf{65}, 045321 (2002)};
	F.~Taddei, M.~Governale, and R.~Fazio, 
	\href {http://dx.doi.org/10.1103/PhysRevB.70.052510}{{\it ibid.} \textbf{70}, 052510 (2004)}.
	
\bibitem{Akhmerov2011}
	A.R.~Akhmerov, J.P.~Dahlhaus, F.~Hassler, M.~Wimmer, and C.W.J.~Beenakker,
	\href{http://prl.aps.org/abstract/PRL/v106/i5/e057001}{\prl~\textbf{106}, 057001 (2011)}.

\bibitem{Wimmer2011}
	M.~Wimmer, A.R.~Akhmerov, J.P.~Dahlhaus, and C.W.J.~Beenakker, 
	\href{http://stacks.iop.org/1367-2630/13/i=5/a=053016}{New J.~Phys.~\textbf{13}, 053016 (2011)}.

\bibitem{Beri2009}
	B.~B\'eri, \href{http://link.aps.org/doi/10.1103/PhysRevB.79.245315}{\prb~\textbf{79}, 245315 (2009)}.

\bibitem{Sanvito1999}
	S.~Sanvito, C.J.~Lambert, J.H.~Jefferson, and A.M.~Bratkovsky,
	\href {http://dx.doi.org/10.1103/PhysRevB.59.11936}{\prb~\textbf{59}, 11936 (1999)}.

\bibitem{Zwierzycki2008}
	M.~Zwierzycki, P.A.~Khomyakov, A.A.~Starikov, K.~Xia, M.~Talanana, P.X.~Xu, V.M.~Karpan, I.~Marushchenko, 
	I.~Turek, G.E.W.~Bauer, G.~Brocks, and P.J.~Kelly,
	\href {http://dx.doi.org/10.1002/pssb.200743359} { Physica Status Solidi B \textbf{245}, 623 (2008)}.
	
\bibitem{quantization}
	Of course $n$ can be any integer, including zero. Hereafter, we shall term the case $Q = n e$ as quantized pumping even when $n=0$. Indeed, the case $Q=0$ is also interesting, since the vanishing of the pumped charge is topologically protected and is not accidental.

\bibitem{SM} See \hyperref[Supplemental Material]{Supplemental Material\ref{app:SM}} for more details. 

\bibitem{Buttiker1990}
	{M.~B\"uttiker}, \href {http://dx.doi.org/10.1103/PhysRevB.41.7906}{ \prb~\textbf{41}, 7906 (1990)}. 

\end{thebibliography}

\begin{thebibliography}{77}	

\bibitem{Law2009}
	K.T.~Law, P.A.~Lee, and T.K.~Ng,
	\href{http://prl.aps.org/abstract/PRL/v103/i23/e237001}{\prl~\textbf{103}, 237001 (2009)}.
	
\bibitem{Flensberg2010}
	K.~Flensberg,
	\href{http://link.aps.org/doi/10.1103/PhysRevB.82.180516}{\prb~\textbf{82}, 180516 (2010)}.

\bibitem{Beenakker1992}
C.W.J.~Beenakker,
\href{http://dx.doi.org/10.1103/PhysRevB.46.12841}{\prb {\bf46}, 12841 (1992)}.

\bibitem{cleanNS}
For a clean NS boundary  normal reflection can be neglected and $s_{\rm A}$ reads  
\begin{equation}
s_{\rm A} = \begin{pmatrix}
0 & r^*_{\rm A}\\
r_{\rm A} & 0
\end{pmatrix}~,
\end{equation} 
%
so that $A = \lsp \openone -r_{\rm A}\, r\, r_{\rm A}^*\, r^*\rsp^{-1} r_{\rm A}$ and
$B = r_{\rm A}^*\, r^*\, r_{\rm A} \lsp \openone -r\, r_{\rm A}^*\, r^*\, r_{\rm A}\rsp^{-1} $.

\bibitem{firstfactor}
This means that $\psi_{n}=0$, $\forall~n$ is not allowed and thus the first factor in Eq.~\eqref{eq:condition} must be non-zero.

\end{thebibliography}
\end{document}